\documentclass[a4paper,conference]{IEEEtran}
\IEEEoverridecommandlockouts
\usepackage[left=1.57cm,right=1.57cm,top=0.95cm,bottom=2.54cm]{geometry}
\usepackage{cite}
\usepackage{amsmath,amssymb,amsfonts}
\usepackage{algorithmic}
\usepackage{graphicx}
\usepackage{textcomp}
\usepackage{xcolor}
\usepackage{subfig}
 \usepackage{multirow}
 \usepackage{pgf}
 \usepackage{pgfplots}
 \usepackage{balance}
 \pgfplotsset{compat=1.18}

\def\mathdefault#1{#1} 

\def\BibTeX{{\rm B\kern-.05em{\sc i\kern-.025em b}\kern-.08em
    T\kern-.1667em\lower.7ex\hbox{E}\kern-.125emX}}

\newcommand\copyrighttext{%
  \footnotesize \textcopyright 2025 IEEE.  Personal use of this material is permitted.  Permission from IEEE must be obtained for all other uses, in any current or future media, including reprinting/republishing this material for advertising or promotional purposes, creating new collective works, for resale or redistribution to servers or lists, or reuse of any copyrighted component of this work in other works.}
\newcommand\copyrightnotice{%
\begin{tikzpicture}[remember picture,overlay]
\node[anchor=south,yshift=10pt] at (current page.south) {\fbox{\parbox{\dimexpr\textwidth-\fboxsep-\fboxrule\relax}{\copyrighttext}}};
\end{tikzpicture}%
}

\begin{document}

\title{Static and Repeated Cooperative Games for the Optimization of the AoI in IoT Networks
\thanks{This work was supported by the European Union under the Italian National Recovery and Resilience Plan (NRRP) Mission 4, Component 2, Investment 1.3, CUP C93C22005250001, partnership on “Telecommunications of the Future” (PE00000001 - program “RESTART”)
}
}

\author{\IEEEauthorblockN{David E. C. R. Catania\textsuperscript{*}, Alessandro Buratto\textsuperscript{*}, and Giovanni Perin\textsuperscript{\dag,*}}
\IEEEauthorblockA{\textsuperscript{*}Department of Information Engineering (DEI), University of Padova (Padova, Italy)}
\IEEEauthorblockA{\textsuperscript{\dag}Department of Information Engineering (DII), University of Brescia (Brescia, Italy)}
\texttt{davidemanuelecorradoraphael.catania@studenti.unipd.it,}\\
\texttt{alessandro.buratto.1@phd.unipd.it, giovanni.perin@unibs.it}
}

\maketitle
\copyrightnotice

\begin{abstract}
Wireless sensing and the internet of things (IoT) are nowadays pervasive in 5G and beyond networks, and they are expected to play a crucial role in 6G. However, a centralized optimization of a distributed system is not always possible and cost-efficient. In this paper, we analyze a setting in which two sensors collaboratively update a common server seeking to minimize the age of information (AoI) of the latest sample of a common physical process. We consider a distributed and uncoordinated setting where each sensor lacks information about whether the other decides to update the server. This strategic setting is modeled through game theory (GT) and two games are defined: i) a static game of complete information with an incentive mechanism for cooperation, and ii) a repeated game over a finite horizon where the static game is played at each stage. We perform a mathematical analysis of the static game finding three Nash Equilibria (NEs) in pure strategies and one in mixed strategies. A numerical simulation of the repeated game is also presented and novel and valuable insight into the setting is given thanks to the definition of a new metric, the price of delayed updates (PoDU), which shows that the decentralized solution provides results close to the centralized optimum.
\end{abstract}


\section{Introduction}
The internet of things (IoT) has deeply transformed various sectors by enabling widespread connectivity between smart devices allowing efficient automation and monitoring \cite{zanella2014internet, elijah2018overview}.
In IoT applications, multiple sources must allocate resources, manage data transmission, and efficiently limit energy consumption \cite{banerjee2023age, chen2020age, buratto2025strategic, badia2021age}.
Quite often, multiple IoT nodes must collaborate to achieve a common goal \cite{he2022collaborative, wu2018game, loomba2014information}.
The large number of wireless devices in IoT networks makes it difficult to centrally optimize action policies for each node in a centralized manner, not only requiring increasingly higher computational costs but also requiring specialized communication resources to possibly avoid congestion in the shared channel \cite{badia2022discounted}.
Game theory (GT) has been applied as a mathematical tool in IoT scenarios to optimize the network in a distributed manner, ensuring low computational complexity and avoiding communication overhead between devices \cite{yang2021game, buratto2025strategic}.
One of the most challenging tasks is to ensure that nodes avoid information staleness by sending updates even if this means incurring a communication cost for them \cite{badia2021age}. Age of information (AoI) is a metric that has been proven in the literature to ensure collaboration between sensors even in distributed scenarios \cite{javani2025age, chen2023analysis, yates2021age}.
Recent studies have demonstrated that optimizing AoI leads to improved network performance in decentralized sensor systems, where nodes must coordinate update transmissions without central control \cite{buratto2023game}. For instance, in multi-source networks, age-based scheduling policies can mitigate data staleness by prioritizing updates from sensors experiencing higher AoI, thereby ensuring that critical information remains fresh at fusion centers \cite{moltafet2020age}.
In this paper, we study a communication scenario between two sensors that are monitoring the same physical process. Their task is to send to a centralized server the collected data independently from each other, knowing that an update from the other sensor makes theirs noninformative.
We model the interaction between the sensors using GT, setting the instantaneous AoI as the objective of the optimization. In addition, we include in the modeling a cost factor that the sensors pay whenever they attempt a transmission and we further limit the number of total transmissions that each node can perform in a finite time horizon \cite{buratto2023optimizing}.
First, we solve the GT formulation in a one-shot static setting and enumerate all the pure- and mixed-strategy Nash equilibria (NEs) of the game. We then leverage these results to study a repeated interaction between the sensors and numerically obtain some subgame perfect equilibria (SPE) strategies for chosen values of the system parameters.
We then introduce a new metric, called the price of delayed updates (PoDU), which measures the efficiency of the average AoI for a distributed solution over an optimal one.
As a result, we show that a distributed solution for this system achieves performance closely comparable to optimality for most system parameter configurations, and it is therefore a good candidate for real-world implementations.

The remainder of this paper is subdivided as follows. In Sec.~\ref{sec:related-work} we discuss relevant literature related to our scenario. In Sec.~\ref{sec:system-model} we describe our system model and provide an extensive description of the game-theoretical setting for both the static and dynamic formulations. We further introduce the price of delayed updates metric. In Sect.~\ref{sec:static-sol} we provide a mathematical analysis to obtain the NEs in the static scenario and we discuss the bounds on the parameters of the system for the emergence of such equilibria. In Sec.~\ref{sec:repeated-results} we show numerical solutions to the dynamic repeated game case and we analyze the system performance by means of the PoDU. Finally, Sec.~\ref{sec:conclusions} concludes the paper.

\section{Related Work}\label{sec:related-work}
The problem of AoI optimization in IoT networks has become increasingly important in recent years \cite{han2021age,feng2021age, shi2024optimize}. Many works consider optimal centralized approaches to solve this task. In \cite{javani2025age} the authors derive closed-form expressions for the average AoI and provide algorithms for computing the AoI in a scenario with multiple sources and multiple servers equipped with last-come first-serve queues.
In \cite{chen2023analysis}, the authors consider a dual sender scenario in which they model the transmission process as two separate queueing systems and derive expressions for the average and peak AoI, proving the advantages of separate queueing systems compared to a unified one with multiple servers. Unlike our approach, they consider updates from the two sensors to be uncorrelated in the information content, and the sensors do not perform any strategic interaction with each other.
Similarly, in \cite{moltafet2020age}, the authors consider a communication scenario with multiple sensors updating a single server that maintains a single queue. They provide expressions for the AoI in this scenario and bounds for the traffic rate sustainable by the system, together with optimal rates to minimize the average AoI.

Another research direction focuses on the use of GT for distributed AoI optimization \cite{yang_channel_2021, yang2021game, badia2024game}.
For instance, the authors of \cite{buratto2023game} consider a multi-sensor scenario in which multiple nodes have to actively collaborate to minimize the average AoI at the receiver side. They obtain a distributed solution by solving a static game of complete information, and they prove that there is the need for some incentive mechanism to guarantee multiple simultaneous transmissions.

Our contribution is inspired by the work of \cite{buratto2023optimizing} for scheduling updates for a single sensor over a finite time horizon. In this reference paper, the authors solve the optimization problem in the presence of an external stochastic updater by obtaining a scheduling policy with limited communication attempts over a finite time horizon using backward induction. It is relevant to note that the external updater is not an active part of the optimization problem: its updates to the server, although beneficial to the considered sensor, are not being optimized to lower the average AoI. Moreover, the authors do not consider strategic interactions between the senders.
Another relevant work is \cite{badia2021age}, in which the author considers a double-sender scenario where two sensors send correlated data to a common receiver and try to minimize the average AoI. The author models the scenario as a static game of complete information and analytically obtains the NE. Differently from our contribution, this work does not include an incentive mechanism for the sensors' interaction and does not analyze the evolution of the strategic interaction in a dynamic setting.

\section{System Model}\label{sec:system-model}
\subsection{Setup}
\begin{figure}[tb]
    \centering
    \includegraphics[width=0.7\columnwidth]{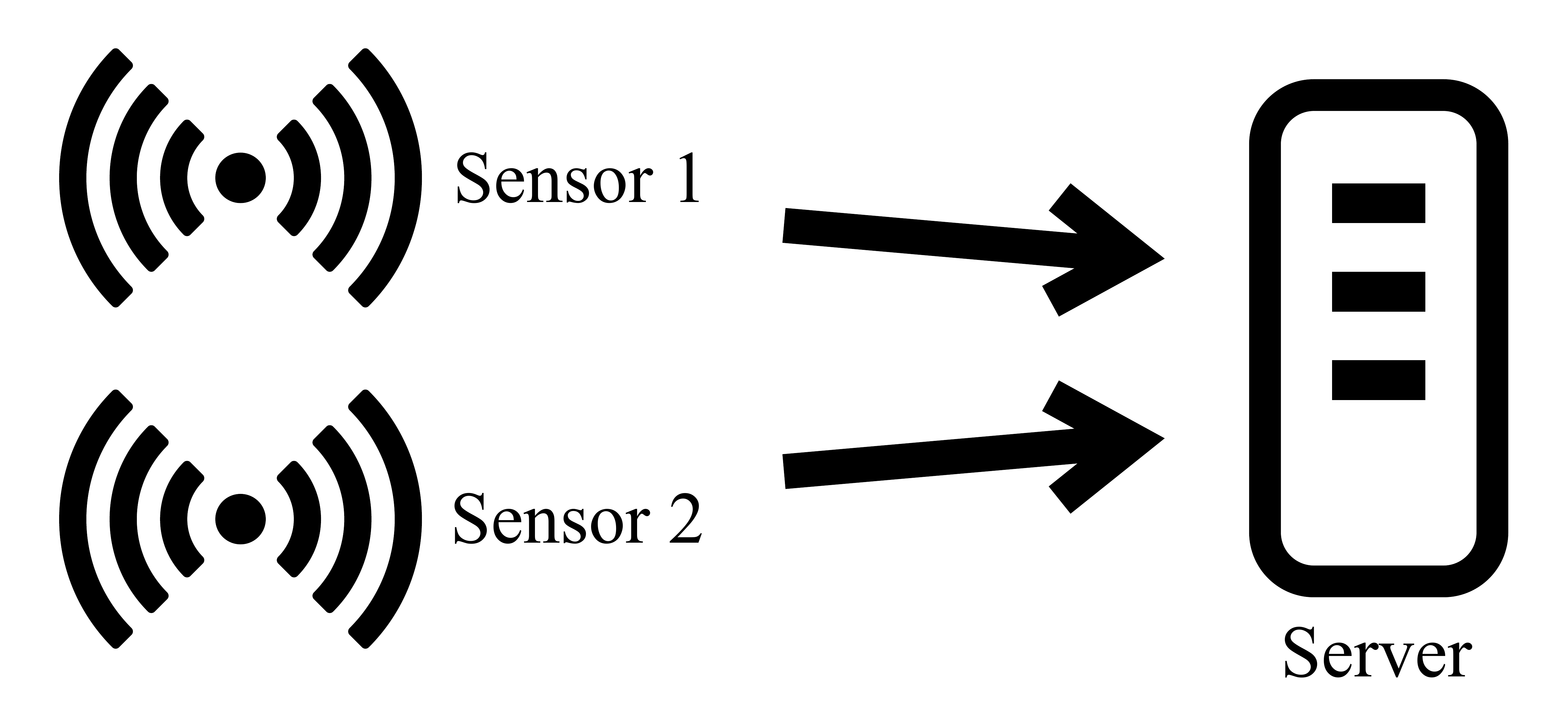}
    \caption{Network configuration}
    \label{fig:setup}
\end{figure}

We consider a collaborative sensing scenario where the server, which will be referred to as the \emph{receiver} for the remainder of this paper, aggregates data collected from two wireless sensors $i\in\mathcal{N}=\{1,2\}$. The latter collect data from the environment according to a physical process $\mathcal{P}$ common to all sensors. A graphical representation of this scenario is shown in Fig.~\ref{fig:setup}.
To track the \emph{freshness} of the information at the receiver's side we employ the age of information metric $\delta(t)$~\cite{yates2021age}, formally defined as the difference between the current time instant $t$ and the timestamp of the latest received update $\tau_k$:
\begin{equation} \label{eq:aoi}
    \delta(t) = t - \tau_k .
\end{equation}

For our analysis, we consider time to be divided into slots, all having the same duration, and a finite time horizon $T$ over which sensors schedule a limited number of updates $G_i \in \mathbb{N}$ with $\sum_i G_i \ll T$. This limit on the total number of updates can encode multiple factors, such as the limited energy available at the transmitter~\cite{hao2019robust, buratto2023optimizing} -- a scenario especially fitted to devices residing in hard-to-reach locations and for which resource management is a challenging task. An alternative physical meaning of tokens can be found in works considering energy-harvesting devices, for which they represent their ability to gather sufficient energy for transmissions~\cite{badia2022discounted, zheng2019closed, shi2024optimize}. We also consider that sensors incur a cost each time they attempt a transmission as is commonly done in the literature~\cite{buratto2023game, hamrouni2024aoi, buratto2025strategic}.


\subsection{Static Game Definition}
We define a static game of complete information to study the equilibria of the system in a one-shot interaction.
To this end, the game $\mathcal{G} = \{\mathcal{N}, \mathcal{S}, \mathcal{U}\}$ is defined, where $\mathcal{N}$ is the set of players in the game; in our case the two sensors, as the server is a passive element of the network. $\mathcal{S} = \{S_1, S_2\}$ is the set of players' strategies: for Player $i$, the possible strategies are $S_i=\{0,1\}$ where $s_i=1$ and $s_i=0$ correspond to the sensor deciding to transmit or not, respectively. We further introduce $s_{-i}$ to refer to the strategy chosen by all the other players in the game, in our scenario the only other player. Finally, $\mathcal{U}=\{u_1, u_2\}$ is the set of utility functions of the players, encoding preferences.

\subsection{Utility function}
Since our aim is to monitor the AoI at the receiver end, each player must evaluate the choice of its strategy and that of the other player based on their impact on the AoI. For the scenario considered, the AoI is reset whenever either sensor updates the server; otherwise, it increases linearly.
We further introduce two additive terms to the utility to deepen the analysis on the AoI in a more realistic scenario.
The first is a cost factor $c_i \in \mathbb{R}^{+}$ that a player pays when performing a transmission. This term may have different values for each player and may include energy consumption for the transmission itself as well as \emph{intangible} costs such as the privacy of the exchanged data \cite{buratto2023game}. The second term is an incentive for the player to keep as many future transmission opportunities as possible. To model this effect, we choose to add a logarithmic function of the current number of tokens $G_i$ available to the player. The choice of the logarithm is also motivated by the fact that we want to capture diminishing returns for players when saving \emph{too many} tokens. We also weight this logarithmic term by a coefficient $\alpha_i \in \mathbb{R}^{+}$. With these considerations, the utility function that the players ought to maximize is given by
\begin{IEEEeqnarray}{rCl}
    u_i(s_i,\, s_{-i}) =& -&\delta(t) (1 - s_i)(1 - s_{-i}) - c_i s_i+ \nonumber \\
    &+& \alpha_i \ln(1 + G_i - s_i) .
\end{IEEEeqnarray}

Note that the negative sign for the AoI $\delta(t)$ and the cost term $c_i s_i$ encodes the fact that players want to simultaneously minimize both the AoI and their own communication cost, whereas the logarithmic incentive is positive as it is the only term that players want to maximize.

With the given definitions, the game $\mathcal{G}$ can be written in normal form as a bi-matrix composed of the payoffs obtained by the players for each combination of their strategies. A graphical representation is reported in Tab.~\ref{tab:matrix}.
\begin{table}[tb]
    \centering
    \begin{equation*}
        \multirow{1}{*}{\rotatebox[origin=c]{90}{\textit{Player 1}\hspace{-.3cm}}}
        \hspace{1mm}
        \begin{array}{c|c|c}
            \multicolumn{3}{c}{\textit{Player 2}} \\
             & \textbf{1} & \textbf{0} \\
             \hline
             \textbf{1} & \begin{array}{c} -c_1 + \alpha_1  \ln(G_1) \\ -c_2 + \alpha_2  \ln(G_2) \end{array}
              & \begin{array}{c} -c_1 + \alpha_1  \ln(G_1) \\ \alpha_2  \ln(G_2 + 1) \end{array} \\
            \hline
            \textbf{0} & \begin{array}{c} \alpha_1  \ln(G_1 + 1) \\ -c_2 + \alpha_2  \ln(G_2) \end{array}
              & \begin{array}{c} -\delta(t) + \alpha_1  \ln(G_1 + 1) \\ -\delta(t) + \alpha_2  \ln(G_2 + 1) \end{array} \\ 
        \end{array}
    \end{equation*}
    \caption{Normal form of the game $\mathcal{G}$.}
    \label{tab:matrix}
\end{table}

\subsection{Dynamic Game Formulation}
\label{sec:dyn_game}
We formulate a dynamic version of the interaction between the sensors as a repeated version of the static game over a finite time horizon $T$. Consequently, we consider time divided into slots of equal length during which the sensors play for the same utilities described previously. We do not include a discount factor for the utility of future rounds.

In this scenario, we are interested in finding the subgame perfect equilibria (SPE) of the game, a sequence of strategies for each player in which they play an NE at every round. To evaluate the performance of a distributed solution, we are interested in the time average of the instantaneous AoI $\delta(t)$ computed as 
\begin{equation}
    \mathbb{E}[\delta(t)] = \frac{1}{T}\sum_{t=0}^T \delta(t) \, .
\end{equation}

\subsection{Price of Delayed Updates}
To better describe the efficiency of our distributed solution with respect to a centralized optimal one in which the updates are scheduled at constant time intervals and the sensors never transmit at the same time, we introduce the price of delayed updates (PoDU), which is computed as the ratio between the average AoI of the distributed solution $\mathbb{E}\left[\delta^{\rm NE}(t)\right]$ and the average AoI of the optimal centralized one $\mathbb{E}\left[\delta^{\rm OPT}(t)\right]$. Formally
\begin{equation}
    {\rm PoDU} = \frac{\mathbb{E}\left[\delta^{\rm NE}(t)\right]}{\mathbb{E}\left[\delta^{\rm OPT}(t)\right]} \, .
\end{equation}
The introduction of this metric is motivated by the fact that strategic sensors may delay the transmissions beyond the optimal time instant because saving resources for future communications attempts gives them higher utility.
This metric can take values in the range $[1,+\infty)$, where $1$ means that the distributed solution is equivalent to the optimal one. Instead, if values approach infinity, it indicates that a distributed solution needs more accurate incentives in order to be a viable solution to the problem at hand.

\section{Static Game Solution}\label{sec:static-sol}
\subsection{Mathematical Analysis}
In this section, we perform an analysis of the static game described previously.
We begin by finding the pure-strategy Nash equilibria of the game.
The examination of the normal form in Tab. \ref{tab:matrix} reveals a significant strategic insight: when the opposing Player $ -i $ transmits, Player $ i $ is discouraged from doing so. This is formally expressed by the condition $ u_i(0,\,1) > u_i(1,\,1) $, which results in
\begin{equation}
    \alpha_i \ln(G_i + 1) > -c_i + \alpha_i \ln(G_i) \quad \forall c_i, \alpha_i
\end{equation}

In contrast, when the opponent refrains from transmitting, Player $ i $ must decide whether to transmit based solely on the trade-off between updating the central computer and conserving resources. In this scenario, Player $ i $ will choose to transmit if $ u_i(1,\,0) > u_i(0,\,0) $, resulting in
\begin{equation}
    -c_i + \alpha_i \ln(G_i) > -\delta(t) + \alpha_i \ln(G_i + 1)\,.
\end{equation}
The solution with respect to $\delta(t)$ provides a closed-form expression for the threshold behavior regarding the AoI, as
\begin{equation}\label{eq:aoiThresh}
    \theta_i \triangleq c_i + \alpha_i \ln\!\left( \frac{G_i+1}{G_i} \right) \,,
\end{equation}
where $\theta_i$ is the threshold value.

Given the expression for the threshold $\theta_i$, the pure strategy best response (BR) yields the optimal strategy for Player $ i $ based on the action taken by the opposing player, indicated by $ s_{-i} $. Specifically, if the opponent transmits, then the optimal strategy for Player $ i $ is to refrain from transmitting. Conversely, if the opponent does not transmit, Player $ i $ should choose to transmit if the AoI is over the threshold; otherwise, if the AoI is below the threshold, Player $ i $ should not transmit. In the case where the AoI is exactly at the threshold value, the player is indifferent between transmitting and not transmitting. Formally, the BR of Player $ i $ is defined as
\begin{equation}\label{eq:br}
\mathrm{BR}_i(s_{-i}) =
    \begin{cases}
        0 & \quad \text{if } s_{-i} = 1 \\
        1 & \quad \text{if } s_{-i} = 0 \, \wedge \, \delta(t) > \theta_i \\
        0 & \quad \text{if } s_{-i} = 0 \, \wedge \, \delta(t) < \theta_i \\
    \end{cases}\,.
\end{equation}

When players have the ability to transmit simultaneously, that is, $ \delta(t) = \theta_i + \varepsilon_i\,, \quad \forall i $ with $ \varepsilon_i \geq 0 $, the optimal strategy may no longer be purely deterministic. In these situations, players can adopt a mixed strategy, transmitting with a certain probability $ p_i $ to maximize their expected utility.
\begin{figure}[tb]
    \centering
    \input{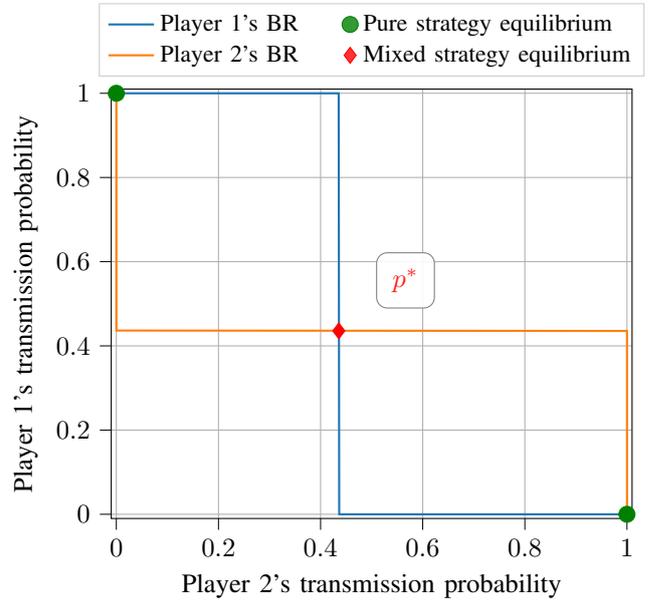}
    \caption{Pure and mixed strategy equilibria where at least one player transmits.}
    \label{fig:nash_eq}
\end{figure}

Let $ p_i $ denote the probability that Player $ i $ transmits, so their mixed strategy is represented as $ \sigma_i(1) = p_i $ and $ \sigma_i(0) = 1 - p_i $. The optimal probability $ p_i $ is determined by solving the following indifference equation, which equates the average utilities of transmitting and not transmitting:
\begin{equation} \label{eq:indifference}
    \mathbb{E}[u_i(1,\, \sigma_{-i})] = \mathbb{E}[u_i(0,\, \sigma_{-i})]\,.
\end{equation}
Specifically, the average utilities are explicitly defined by
\begin{equation}\label{eq:exsys}
    \!\!\begin{cases}
        \mathbb{E}[u_i(1,\, \sigma_{-i})] = -c_i + \alpha_i \ln(G_i) \\
        \begin{aligned}
            \mathbb{E}[u_i(0,\, \sigma_{-i})] ={}& p_{-i} \left[ \alpha_i  \ln(G_i + 1) \right] + \\ 
            &\!( 1 - p_{-i} )  \left[ - \delta(t) + \alpha_i  \ln(G_i + 1) \right]
        \end{aligned}
    \end{cases}\,.
\end{equation}
Substituting $ \delta(t) = \theta_i + \varepsilon_i $ into \eqref{eq:exsys} and solving \eqref{eq:indifference} for $ p_{-i} $ yields
\begin{equation} \label{eq:pi}
    p_{-i}=\varepsilon_i \cdot \left(\frac{1}{ c_i + \alpha_i \ln\!\left( \frac{G_i+1}{G_i} \right) } + 1 \right) \, .
\end{equation}

A similar expression can be found for the probability of indifference of the other Player $p_i$ by swapping the pedices $i$ and $-i$ in \eqref{eq:pi}.
Thus, the mixed strategy equilibrium is for the players to transmit with probability equal to $p_i$ and $p_{-i}$, respectively.
Fig.~\ref{fig:nash_eq} shows a graphical representation of the pure strategy equilibria where only one of the players transmits, and the mixed equilibrium where the players transmit with their own probability in the form of \eqref{eq:pi} where all the parameters $\varepsilon_i$, $c_i$, $\alpha_i$ and $G_i$ are chosen to be the same for both players.



Summing up the previous results, we have shown the existence of three NEs in pure strategies and one in mixed strategies, depending on the relation between AoI and each player's transmission threshold. In pure strategy equilibria, when the AoI remains below both players' thresholds ($\delta(t) < \theta_i$), neither has an incentive to transmit, leading to the NE $ (s_i^*,\,s_{-i}^*) = (0,\,0) $. If the AoI exceeds the threshold for only one player, the NE results in that player transmitting while the other abstains, yielding either $ (s_i^*,\,s_{-i}^*) = (1,\,0) $ or $ (s_i^*,\,s_{-i}^*) = (0,\,1) $. However, when both players reach their transmission threshold at the same time, i.e., $ \delta(t) = \theta_i + \varepsilon_i \quad \forall i$, a single NE emerges in mixed strategies, where each player transmits with probability $ p_i $ as determined by \eqref{eq:pi}. This leads to the equilibrium $ (\sigma_i^*,\,\sigma_{-i}^*) = (p_i,\,p_{-i}) $, ensuring a balance between maintaining fresh information and conserving transmission resources.

\subsection{NEs Discussion}

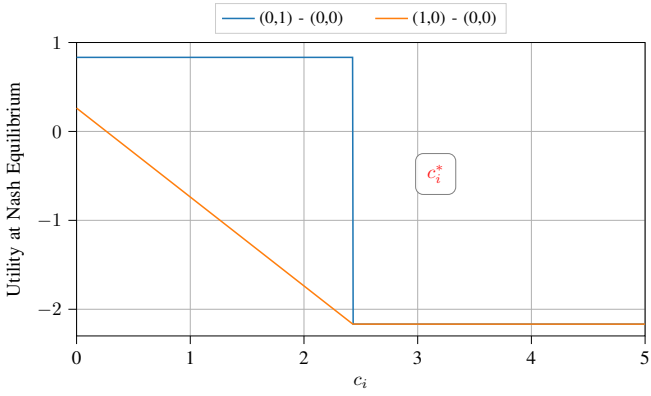
\begin{figure}[tb]
    \centering
    \resizebox{\columnwidth}{!}{
\begin{tikzpicture}

\definecolor{darkgray176}{RGB}{176,176,176}
\definecolor{darkorange25512714}{RGB}{255,127,14}
\definecolor{gray}{RGB}{128,128,128}
\definecolor{lightgray204}{RGB}{204,204,204}
\definecolor{steelblue31119180}{RGB}{31,119,180}

\begin{axis}[
legend cell align={left},
legend style={
    fill opacity=0.8, 
    draw opacity=1, 
    text opacity=1, 
    draw=lightgray204,
    legend columns=2,
    /tikz/every even column/.append style={column sep=0.5cm},
    font=\small,
    at={(0.5,1.03)}, 
    anchor=south
},
width=12cm,
height=7cm,
tick align=outside,
tick pos=left,
x grid style={darkgray176},
xlabel={\(\displaystyle c_i\)},
xmajorgrids,
xmin=0, xmax=5,
xtick style={color=black},
xtick={0,1,2,3,4,5},
xticklabels={0,1,2,3,4,5},
y grid style={darkgray176},
ylabel={Utility at Nash Equilibrium},
ymajorgrids,
ymin=-2.3, ymax=1,
ytick style={color=black}
]
\addplot [thick, steelblue31119180]
table {%
0 0.832909107208252
2.4274275302887 0.832909107208252
2.4324324131012 -2.16709089279175
5 -2.16709089279175
};
\addlegendentry{(0,1) - (0,0)}
\addplot [thick, darkorange25512714]
table {%
0 0.262364268302917
2.4274275302887 -2.16506314277649
2.4324324131012 -2.16709089279175
5 -2.16709089279175
};
\addlegendentry{(1,0) - (0,0)}
\draw (axis cs:3.1582916839921,-0.524728528934982) node[
  scale=1,
  fill=white,
  draw=gray,
  line width=0.4pt,
  inner sep=6pt,
  fill opacity=0.9,
  rounded corners,
  anchor=base,
  text=red,
  rotate=0.0
]{$c_i^*$};
\end{axis}

\end{tikzpicture}}
    \caption{Variation of equilibrium utility with respect to $ c_i $.}
    \label{fig:utility_c}
\end{figure}
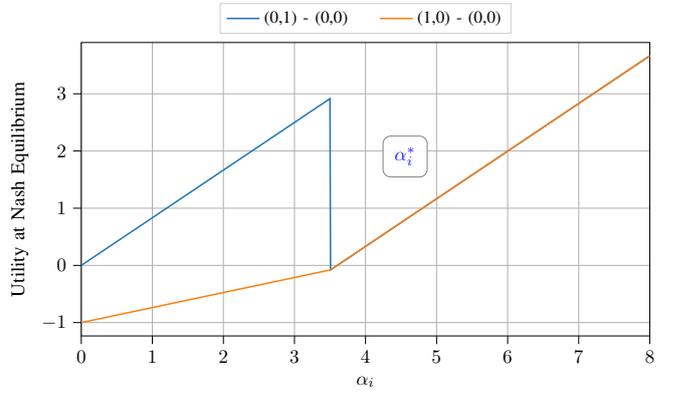
\begin{figure}[tb]
    \centering
    \resizebox{\columnwidth}{!}{
\begin{tikzpicture}

\definecolor{darkgray176}{RGB}{176,176,176}
\definecolor{darkorange25512714}{RGB}{255,127,14}
\definecolor{gray}{RGB}{128,128,128}
\definecolor{lightgray204}{RGB}{204,204,204}
\definecolor{steelblue31119180}{RGB}{31,119,180}

\begin{axis}[
legend cell align={left},
legend style={
    fill opacity=0.8, 
    draw opacity=1, 
    text opacity=1, 
    draw=lightgray204,
    legend columns=2,
    /tikz/every even column/.append style={column sep=0.5cm},
    font=\small,
    at={(0.5,1.03)}, 
    anchor=south
},
tick align=outside,
tick pos=left,
width=12cm,
height=7cm,
x grid style={darkgray176},
xlabel={\(\displaystyle \alpha_i\)},
xmajorgrids,
xmin=0, xmax=8,
xtick style={color=black},
y grid style={darkgray176},
ylabel={Utility at Nash Equilibrium},
ymajorgrids,
ymin=-1.23316364917404, ymax=3.89643663265487,
ytick style={color=black}
]
\addplot [thick, steelblue31119180]
table {%
0 0
3.49949955940247 2.91476511955261
3.50750756263733 -0.0785650014877319
8 3.66327309608459
};
\addlegendentry{(0,1) - (0,0)}
\addplot [thick, darkorange25512714]
table {%
0 -1
3.50750756263733 -0.0785650014877319
8 3.66327309608459
};
\addlegendentry{(1,0) - (0,0)}
\draw (axis cs:4.55704746333735,1.83163649174042) node[
  scale=1,
  fill=white,
  draw=gray,
  line width=0.4pt,
  inner sep=6pt,
  fill opacity=0.9,
  rounded corners,
  anchor=base,
  text=blue,
  rotate=0.0
]{$\alpha_i^*$};
\end{axis}

\end{tikzpicture}}
    \caption{Variation of equilibrium utility with respect to $ \alpha_i $.}
    \label{fig:utility_a}
\end{figure}

The decision threshold \eqref{eq:aoiThresh} reveals two key effects that shape the strategic behavior of the players. First, an increase in transmission cost $ c_i $ raises the threshold $ \theta_i $, indicating that the player is willing to tolerate a higher AoI before transmitting. This leads to less frequent updates and, consequently, a higher average AoI. Second, the parameter $ \alpha_i $ influences the weight assigned to resource conservation. A higher $ \alpha_i $ increases the incentive to save tokens, causing the player to delay transmission; thus, even a moderate transmission cost can lead to a sufficiently high $ \theta_i $ to impact the decision-making process.  

To further examine these dependencies, each parameter can be isolated, yielding the expressions
\begin{equation}
\begin{aligned}
     c_i^* &= \delta(t) - \alpha_i \ln\!\left( \frac{G_i+1}{G_i} \right)\,, \quad {\rm and}\\
     \alpha_i^* &= \frac{\delta(t) - c_i}{\ln\!\left( \frac{G_i+1}{G_i} \right)}\,.
\end{aligned}
\end{equation}
These values identify critical thresholds beyond which the strategic equilibrium changes. Specifically, when $ c_i > c_i^* $, the cost outweighs the benefit of transmission, causing the player to refrain from sending updates. Similarly, if $ \alpha_i > \alpha_i^* $, saving resources is a high priority for the player.  

Figures~\ref{fig:utility_c} and \ref{fig:utility_a} highlight how variations in $ c_i $ and $ \alpha_i $ influence the equilibrium utility. The results reveal distinct transitions between regions where the equilibrium favors asymmetric strategies -- such as $ (0,1) $ or $ (1,0) $ -- and regions where neither player transmits ($ (0,0) $). These findings emphasize the sensibility of the equilibria to system parameters as $c_i$ and $\alpha_i$ closely control the stability and efficiency of a distributed solution.

\begin{figure*}[tb]
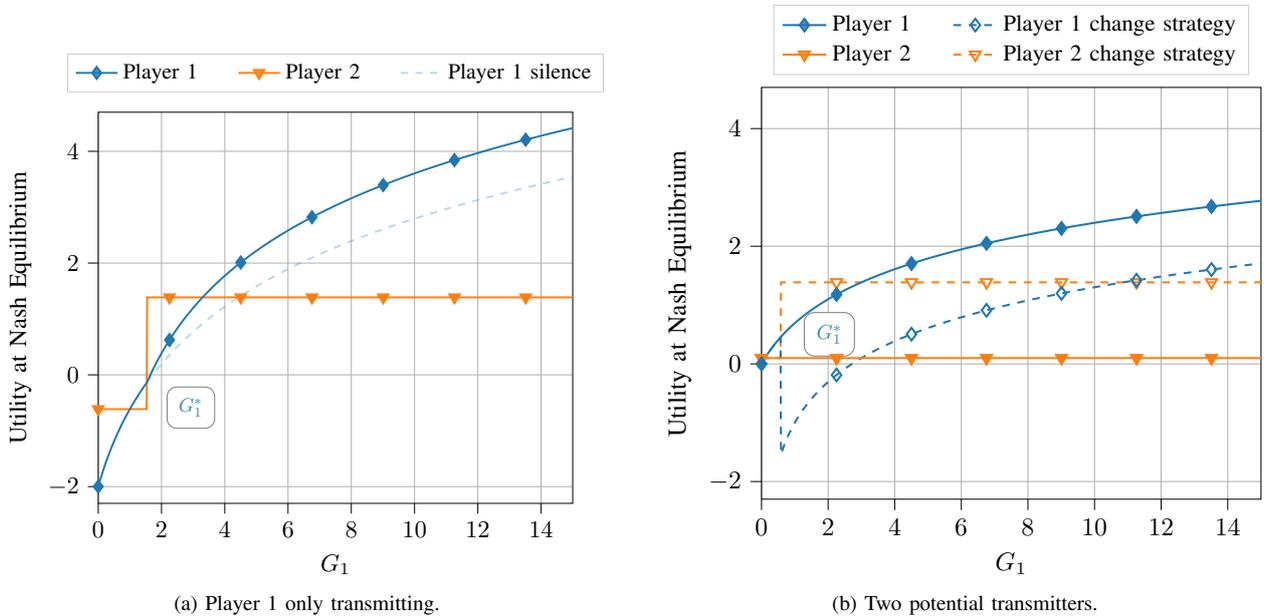

    \centering
    \subfloat[Player 1 only transmitting.\label{fig:utility_g_NN_TN}]{\resizebox{0.45\textwidth}{!}{\input{figures/utility_g_NN_TN}}}
    \hspace{.5cm}
    \subfloat[Two potential transmitters.\label{fig:utility_g_TN_NT}]{\resizebox{0.45\textwidth}{!}{\input{figures/utility_g_TN_NT}}}   
    \caption{Utility at equilibrium as a function of $G_1$.}
    \label{fig:utility_G1}
\end{figure*}


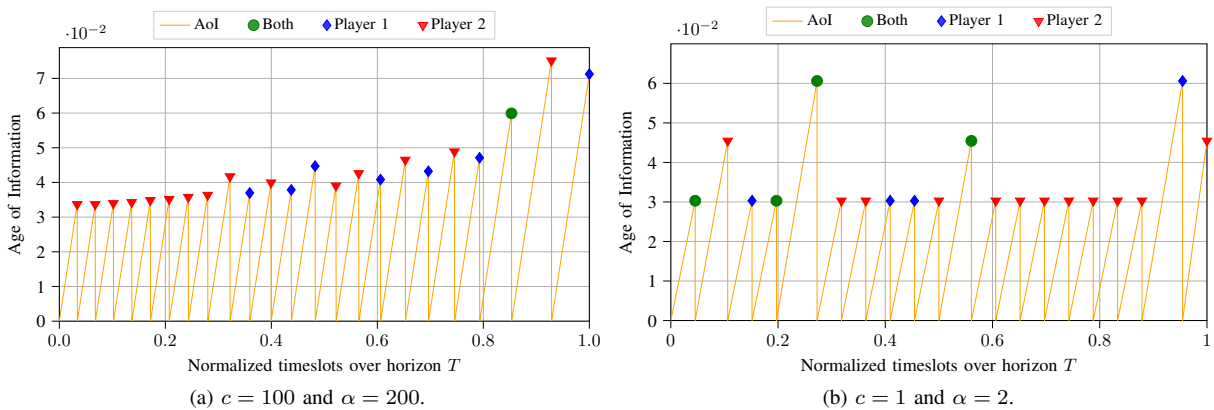
\begin{figure*}[tb]
    \centering
    \subfloat[$ c = 100 $ and $ \alpha = 200 $.\label{fig:repgame_high}]{\resizebox{.45\textwidth}{!}{
\begin{tikzpicture}

\definecolor{darkgray176}{RGB}{176,176,176}
\definecolor{green}{RGB}{0,128,0}
\definecolor{lightgray204}{RGB}{204,204,204}
\definecolor{orange}{RGB}{255,165,0}

\begin{axis}[
  legend cell align={left},
  legend style={
    fill opacity=0.8,
    draw opacity=1,
    text opacity=1,
    draw=lightgray204,
    legend columns=4,
    /tikz/every even column/.append style={column sep=0.5cm},
    font=\small,
    at={(0.5,1.03)}, 
    anchor=south
  },
width=12cm,
height=7cm,
minor xtick={},
minor ytick={},
tick align=outside,
tick pos=left,
x grid style={darkgray176},
xlabel={Normalized timeslots over horizon \(\displaystyle T\)},
xmajorgrids,
xmin=0, xmax=1,
xtick style={color=black},
xtick={0,0.2,0.4,0.6,0.8,1},
xticklabels={
  \(\displaystyle {0.0}\),
  \(\displaystyle {0.2}\),
  \(\displaystyle {0.4}\),
  \(\displaystyle {0.6}\),
  \(\displaystyle {0.8}\),
  \(\displaystyle {1.0}\)
},
y grid style={darkgray176},
ylabel={Age of Information},
ymajorgrids,
ymin=0, ymax=0.0788673621460507,
ytick style={color=black},
ytick={0,0.01,0.02,0.03,0.04,0.05,0.06,0.07},
yticklabels={
  0,1,2,3,4,5,6,7
}
]
\addplot [line width=0.52pt, orange]
table {%
0 0
0.0336810730253353 0.0336810730253353
0.0339791356184799 0
0.0676602086438152 0.0336810730253353
0.0679582712369598 0
0.10193740685544 0.0339791356184799
0.102235469448584 0
0.136512667660209 0.0342771982116244
0.136810730253353 0
0.171684053651267 0.0348733233979136
0.171982116244411 0
0.207153502235469 0.0351713859910581
0.207451564828614 0
0.243219076005961 0.0357675111773472
0.243517138599106 0
0.279880774962742 0.0363636363636364
0.280178837555887 0
0.321907600596125 0.0417287630402385
0.32220566318927 0
0.359165424739195 0.0369597615499255
0.35946348733234 0
0.399403874813711 0.0399403874813711
0.399701937406855 0
0.437555886736215 0.0378539493293592
0.437853949329359 0
0.482563338301043 0.0447093889716841
0.482861400894188 0
0.521907600596125 0.0390461997019374
0.52220566318927 0
0.564828614008942 0.0426229508196721
0.565126676602086 0
0.605961251862891 0.0408345752608048
0.606259314456036 0
0.652757078986587 0.0464977645305514
0.653055141579732 0
0.696274217585693 0.0432190760059613
0.696572280178838 0
0.745454545454545 0.0488822652757079
0.74575260804769 0
0.792846497764531 0.0470938897168405
0.793144560357675 0
0.853055141579732 0.0599105812220566
0.853353204172876 0
0.928464977645306 0.0751117734724292
0.92876304023845 0
1 0.0712369597615499
};
\addlegendentry{AoI}
\addplot [only marks, draw=green, mark size=3, draw=none, fill=green, mark=*]
table{%
0.853055141579732 0.0599105812220566
};
\addlegendentry{Both}
\addplot [only marks, blue, mark=diamond*, mark size=3, mark options={solid}]
table{%
0.359165424739195 0.0369597615499255
};
\addlegendentry{Player 1}
\addplot [semithick, red, mark=triangle*, mark size=3, mark options={solid,rotate=180}, only marks]
table{%
0.0336810730253353 0.0336810730253353
};
\addlegendentry{Player 2}
\addplot [semithick, red, mark=triangle*, mark size=3, mark options={solid,rotate=180}, forget plot]
table{%
0.0676602086438152 0.0336810730253353
};
\addplot [semithick, red, mark=triangle*, mark size=3, mark options={solid,rotate=180}, forget plot]
table{%
0.10193740685544 0.0339791356184799
};
\addplot [semithick, red, mark=triangle*, mark size=3, mark options={solid,rotate=180}, ]
table{%
0.136512667660209 0.0342771982116244
};
\addplot [semithick, red, mark=triangle*, mark size=3, mark options={solid,rotate=180}, ]
table{%
0.171684053651267 0.0348733233979136
};
\addplot [semithick, red, mark=triangle*, mark size=3, mark options={solid,rotate=180}, ]
table{%
0.207153502235469 0.0351713859910581
};
\addplot [semithick, red, mark=triangle*, mark size=3, mark options={solid,rotate=180}, ]
table{%
0.243219076005961 0.0357675111773472
};
\addplot [semithick, red, mark=triangle*, mark size=3, mark options={solid,rotate=180}, ]
table{%
0.279880774962742 0.0363636363636364
};
\addplot [semithick, red, mark=triangle*, mark size=3, mark options={solid,rotate=180}, ]
table{%
0.321907600596125 0.0417287630402385
};
\addplot [semithick, red, mark=triangle*, mark size=3, mark options={solid,rotate=180}, ]
table{%
0.399403874813711 0.0399403874813711
};
\addplot [only marks, blue, mark=diamond*, mark size=3, mark options={solid}]
table{%
0.437555886736215 0.0378539493293592
};
\addplot [only marks, blue, mark=diamond*, mark size=3, mark options={solid}]
table{%
0.482563338301043 0.0447093889716841
};
\addplot [semithick, red, mark=triangle*, mark size=3, mark options={solid,rotate=180}, ]
table{%
0.521907600596125 0.0390461997019374
};
\addplot [semithick, red, mark=triangle*, mark size=3, mark options={solid,rotate=180}, ]
table{%
0.564828614008942 0.0426229508196721
};
\addplot [only marks, blue, mark=diamond*, mark size=3, mark options={solid}]
table{%
0.605961251862891 0.0408345752608048
};
\addplot [semithick, red, mark=triangle*, mark size=3, mark options={solid,rotate=180}, ]
table{%
0.652757078986587 0.0464977645305514
};
\addplot [only marks, blue, mark=diamond*, mark size=3, mark options={solid}]
table{%
0.696274217585693 0.0432190760059613
};
\addplot [semithick, red, mark=triangle*, mark size=3, mark options={solid,rotate=180}, ]
table{%
0.745454545454545 0.0488822652757079
};
\addplot [only marks, blue, mark=diamond*, mark size=3, mark options={solid}]
table{%
0.792846497764531 0.0470938897168405
};

\addplot [semithick, red, mark=triangle*, mark size=3, mark options={solid,rotate=180}, ]
table{%
0.928464977645306 0.0751117734724292
};
\addplot [only marks, blue, mark=diamond*, mark size=3, mark options={solid}]
table{%
1 0.0712369597615499
};
\end{axis}

\end{tikzpicture}}}
    \subfloat[$ c = 1 $ and $ \alpha = 2 $.\label{fig:repgame_low}]{\resizebox{.45\textwidth}{!}{
\begin{tikzpicture}

\definecolor{darkgray176}{RGB}{176,176,176}
\definecolor{green}{RGB}{0,128,0}
\definecolor{lightgray204}{RGB}{204,204,204}
\definecolor{orange}{RGB}{255,165,0}

\begin{axis}[
legend cell align={left},
legend style={
  fill opacity=0.8,
  draw opacity=1,
  text opacity=1,
  draw=lightgray204,
  legend columns=4,
  /tikz/every even column/.append style={column sep=0.5cm},
  font=\small,
  at={(0.5,1.03)}, 
  anchor=south
},
width=12cm,
height=7cm,
minor xtick={},
minor ytick={},
tick align=outside,
tick pos=left,
x grid style={darkgray176},
xlabel={Normalized timeslots over horizon \(\displaystyle T\)},
xmajorgrids,
xmin=0, xmax=1,
xtick style={color=black},
xtick={0,.2,.4,.6,.8,1},
y grid style={darkgray176},
ylabel={Age of Information},
ymajorgrids,
ymin=0, ymax=0.07,
ytick style={color=black},
ytick={0,.01,.02,.03,.04,.05,.06},
yticklabels={0,1,2,3,4,5,6}
]
\addplot [line width=0.52pt, orange]
table {%
0 0
0.0454545454545455 0.0303030303030303
0.0454545454545455 0
0.106060606060606 0.0454545454545455
0.106060606060606 0
0.151515151515152 0.0303030303030303
0.151515151515152 0
0.196969696969697 0.0303030303030303
0.196969696969697 0
0.272727272727273 0.0606060606060606
0.272727272727273 0
0.318181818181818 0.0303030303030303
0.318181818181818 0
0.363636363636364 0.0303030303030303
0.363636363636364 0
0.409090909090909 0.0303030303030303
0.409090909090909 0
0.454545454545455 0.0303030303030303
0.454545454545455 0
0.5 0.0303030303030303
0.5 0
0.560606060606061 0.0454545454545455
0.560606060606061 0
0.606060606060606 0.0303030303030303
0.606060606060606 0
0.651515151515151 0.0303030303030303
0.651515151515151 0
0.696969696969697 0.0303030303030303
0.696969696969697 0
0.742424242424242 0.0303030303030303
0.742424242424242 0
0.787878787878788 0.0303030303030303
0.787878787878788 0
0.833333333333333 0.0303030303030303
0.833333333333333 0
0.878787878787879 0.0303030303030303
0.878787878787879 0
0.954545454545455 0.0606060606060606
0.954545454545455 0
1 0.0454545454545455
};
\addlegendentry{AoI}
\addplot [only marks, green, mark=*, mark size=3, mark options={solid}]
table{%
0.0454545454545455 0.0303030303030303
};
\addlegendentry{Both}
\addplot [only marks, blue, mark=diamond*, mark size=3, mark options={solid}]
table{%
0.151515151515152 0.0303030303030303
};
\addlegendentry{Player 1}
\addplot [only marks, red, mark=triangle*, mark size=3, mark options={solid,rotate=180}]
table{%
0.106060606060606 0.0454545454545455
};
\addlegendentry{Player 2}
\addplot [only marks, green, mark=*, mark size=3, mark options={solid}]
table{%
0.196969696969697 0.0303030303030303
};
\addplot [only marks, green, mark=*, mark size=3, mark options={solid}]
table{%
0.272727272727273 0.0606060606060606
};
\addplot [only marks, red, mark=triangle*, mark size=3, mark options={solid,rotate=180}]
table{%
0.318181818181818 0.0303030303030303
};
\addplot [only marks, red, mark=triangle*, mark size=3, mark options={solid,rotate=180}]
table{%
0.363636363636364 0.0303030303030303
};
\addplot [only marks, blue, mark=diamond*, mark size=3, mark options={solid}]
table{%
0.409090909090909 0.0303030303030303
};
\addplot [only marks, blue, mark=diamond*, mark size=3, mark options={solid}]
table{%
0.454545454545455 0.0303030303030303
};
\addplot [only marks, red, mark=triangle*, mark size=3, mark options={solid,rotate=180}]
table{%
0.5 0.0303030303030303
};
\addplot [only marks, green, mark=*, mark size=3, mark options={solid}]
table{%
0.560606060606061 0.0454545454545455
};
\addplot [only marks, red, mark=triangle*, mark size=3, mark options={solid,rotate=180}]
table{%
0.606060606060606 0.0303030303030303
};
\addplot [only marks, red, mark=triangle*, mark size=3, mark options={solid,rotate=180}]
table{%
0.651515151515151 0.0303030303030303
};
\addplot [only marks, red, mark=triangle*, mark size=3, mark options={solid,rotate=180}]
table{%
0.696969696969697 0.0303030303030303
};
\addplot [only marks, red, mark=triangle*, mark size=3, mark options={solid,rotate=180}]
table{%
0.742424242424242 0.0303030303030303
};
\addplot [only marks, red, mark=triangle*, mark size=3, mark options={solid,rotate=180}]
table{%
0.787878787878788 0.0303030303030303
};
\addplot [only marks, red, mark=triangle*, mark size=3, mark options={solid,rotate=180}]
table{%
0.833333333333333 0.0303030303030303
};
\addplot [only marks, red, mark=triangle*, mark size=3, mark options={solid,rotate=180}]
table{%
0.878787878787879 0.0303030303030303
};
\addplot [only marks, blue, mark=diamond*, mark size=3, mark options={solid}]
table{%
0.954545454545455 0.0606060606060606
};
\addplot [only marks, red, mark=triangle*, mark size=3, mark options={solid,rotate=180}]
table{%
1 0.0454545454545455
};
\end{axis}

\end{tikzpicture}}}
    \caption{Evolution of the AoI in the repeated game with $G_1=8$ and $G_2=16$.}  
    \label{fig:dynamic_game}
\end{figure*}

The number of tokens $ G_i $ plays a crucial role in shaping the equilibrium utility by modulating the impact of the logarithmic term. A higher $ G_i $ decreases the incremental effect of this term, thereby lowering the threshold $ \theta_i $ and encouraging earlier transmissions. Conversely, when resources are scarce, a low $ G_i $ results in a higher $ \theta_i $, delaying transmission and increasing AoI. Two distinct scenarios illustrate these effects. In the case of a single transmitter -- such as when Player 2 is unable to transmit -- a number of tokens $ G_1 $ beyond a critical threshold $ G_1^* $ improves utility for both players, as Player 1 can transmit, shifting the equilibrium from $ (0,0) $ to $ (1,0) $ (see Fig.~\ref{fig:utility_g_NN_TN}). In the scenario where Player 2 can transmit, two potential equilibria emerge when Player 1's tokens surpass $ G_1^* $: in one, Player 1 remains silent while Player 2 transmits; in the other, represented by dashed lines, the roles are reversed, with Player 1 transmitting and Player 2 refraining (see Fig.~\ref{fig:utility_g_TN_NT}).




\section{Repeated Game Numerical Results}\label{sec:repeated-results}

We present here a numerical simulation of the dynamic setting formulated in Sec.~\ref{sec:dyn_game}. In this simulation, the players play one of the static game's NEs at each round according to the current state of the system. The results show that when the values of $ c_i $ and $ \alpha_i $ are relatively high in relation to AoI, the cost of transmission collisions or overly aggressive transmissions causes players to strategically spread their transmissions over time. This behavior is illustrated in Fig.~\ref{fig:repgame_high}, which depicts the evolution of the AoI over time (in discrete increments of 1) for a configuration with $ c_1 = c_2 = 100 $, $ \alpha_1 = \alpha_2 = 200 $, $ G_1 = 8 $, and $ G_2 = 16 $. The plot highlights the transmission events for each player, showing how the system successfully addresses the collision issue. Specifically, since Player 2 has twice as many transmission tokens as Player 1, at the beginning of the simulation we observe that only Player 2 updates the receiver, while Player 1 keeps the tokens for future transmissions. Only starting from the moment in which both the players have the same amount of tokens (after 8 updates), the two sensors start to divide the updates among each other. The only collision is observed in correspondence of the third-last update, in a region where the AoI was higher because both sensors have few tokens left.

On the other hand, when the values of $ c_i $ and $ \alpha_i $ are closer to the order of magnitude of the time increment, the influence of cost and resource conservation is insufficient to minimize the risk of collisions and maximize the average payoffs. This behavior is demonstrated in Fig. \ref{fig:repgame_low}, where the evolution of AoI shows a less strategic distribution of transmissions. Since the relative weight of the AoI is now higher concerning the transmission cost and the available tokens, both sensors prefer to update the receiver sooner. This results in an increased risk of collisions.

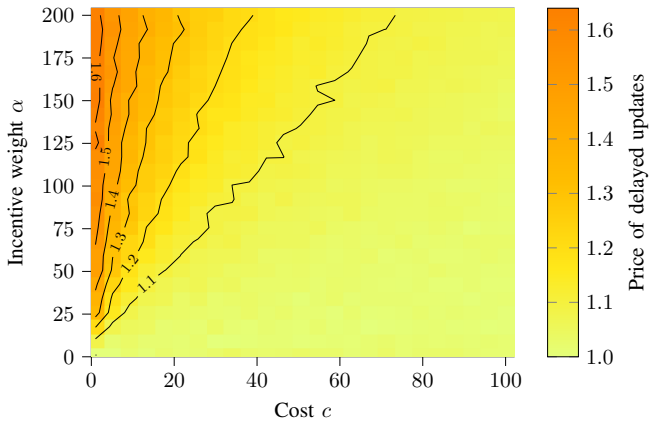
\begin{figure}[tb]
    \centering
    \resizebox{\columnwidth}{!}{
\begin{tikzpicture}

\definecolor{darkgray176}{RGB}{176,176,176}

\begin{axis}[
colorbar,
colorbar style={ytick={1,1.1,1.2,1.3,1.4,1.5,1.6,1.7},yticklabels={
  \(\displaystyle {1.0}\),
  \(\displaystyle {1.1}\),
  \(\displaystyle {1.2}\),
  \(\displaystyle {1.3}\),
  \(\displaystyle {1.4}\),
  \(\displaystyle {1.5}\),
  \(\displaystyle {1.6}\),
  \(\displaystyle {1.7}\)
},ylabel={Price of delayed updates}},
colormap={mymap}{[1pt]
  rgb(0pt)=(0.894117647058824,1,0.47843137254902);
  rgb(1pt)=(1,0.909803921568627,0.101960784313725);
  rgb(2pt)=(1,0.741176470588235,0);
  rgb(3pt)=(1,0.627450980392157,0);
  rgb(4pt)=(0.988235294117647,0.498039215686275,0)
},
point meta max=1.6400006967016,
point meta min=1,
tick align=outside,
tick pos=left,
x grid style={darkgray176},
xlabel={Cost \(\displaystyle c\)},
xmin=0, xmax=102.0625,
xtick style={color=black},
xtick={0,20,40,60,80,100,120},
xticklabels={
  \(\displaystyle {0}\),
  \(\displaystyle {20}\),
  \(\displaystyle {40}\),
  \(\displaystyle {60}\),
  \(\displaystyle {80}\),
  \(\displaystyle {100}\),
  \(\displaystyle {120}\)
},
y grid style={darkgray176},
ylabel={Incentive weight \(\displaystyle \alpha\)},
ymin=0, ymax=204.145833333333,
ytick style={color=black},
ytick={0,25,50,75,100,125,150,175,200,225},
yticklabels={
  \(\displaystyle {0}\),
  \(\displaystyle {25}\),
  \(\displaystyle {50}\),
  \(\displaystyle {75}\),
  \(\displaystyle {100}\),
  \(\displaystyle {125}\),
  \(\displaystyle {150}\),
  \(\displaystyle {175}\),
  \(\displaystyle {200}\),
  \(\displaystyle {225}\)
}
]
\addplot graphics [includegraphics cmd=\pgfimage,xmin=0, xmax=102.0625, ymin=0, ymax=204.145833333333] {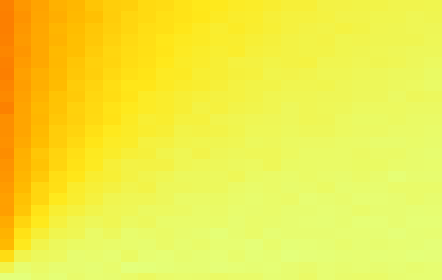};
\addplot [draw=black, draw=black]
table{%
x  y
1 1
1 1
};
\addplot [draw=black, draw=black]
table{%
x  y
29.875 9.29166666666667
29.875 9.29166666666667
29.875 9.29166666666667
29.875 9.29166666666667
29.875 9.29166666666667
};
\addplot [draw=black, draw=black]
table{%
x  y
1 10.5007467275533
4.3792974806571 17.5833333333333
5.125 20.0893128628444
8.03126694663561 25.875
9.25 30.6991345583074
10.8443380296006 34.1666666666667
11.2051956009788 35.3622756282276
};
\addplot [draw=black, draw=black]
table{%
x  y
16.17325051983 47.7328541061373
17.5 50.1595118654671
17.9032818275904 50.75
21.4596391346197 59.0416666666667
21.625 59.8459278031086
24.6323681356187 67.3333333333333
25.75 69.0057489309876
28.386902874106 75.625
27.9126510661264 83.9166666666667
29.875 88.2689789981242
34 90.3612984475239
34.4506788401507 92.2083333333333
34 99.6563692681459
33.9643048233656 100.5
34 100.59118102714
38.125 102.37594118925
40.3973491374401 108.791666666667
42.25 116.41477441376
46.375 116.6594381859
46.4853500938158 117.083333333333
46.375 117.608765330831
44.7538241806455 125.375
46.375 130.164451301617
49.4392388117299 133.666666666667
50.5 135.665139341816
52.8483087603695 141.958333333333
54.625 147.34888006121
58.75 150.232301477552
58.7555945644397 150.25
58.75 150.302032519778
54.625 155.609892522201
54.2351184320596 158.541666666667
54.625 159.041574698908
58.75 161.767490481091
61.9399426433662 166.833333333333
62.875 169.23802452034
64.3717782613752 175.125
66.4144466238377 183.416666666667
67 187.880850959534
70.7295764764975 191.708333333333
71.125 191.850998805191
73.3708826358836 200
};
\addplot [draw=black, draw=black]
table{%
x  y
1 17.3008797546893
1.13476639536402 17.5833333333333
4.10187107200637 25.875
4.77633734825136 34.1666666666667
5.125 37.0034428077003
6.83315790617868 42.4583333333333
7.56947737626926 46.2666021769455
};
\addplot [draw=black, draw=black]
table{%
x  y
11.3554304200854 61.9032100599956
12.5100808571348 67.3333333333333
13.2592309212172 75.625
13.375 76.1849460180435
16.1071897040793 83.9166666666667
16.9027224074693 92.2083333333333
17.5 94.5980345067742
18.8977476364042 100.5
20.513416868675 108.791666666667
21.625 112.971355729398
23.3349966048675 117.083333333333
23.4383529330206 125.375
25.75 133.044897755965
26.017306249762 133.666666666667
25.75 136.765608616088
25.4376132789069 141.958333333333
25.75 143.282497633739
28.177004899766 150.25
29.875 158.524367039921
29.8813359406219 158.541666666667
31.3042383512994 166.833333333333
32.7105525551108 175.125
33.8712265394277 183.416666666667
34 183.976311910163
36.2517752723052 191.708333333333
38.125 197.119406207958
39.0155121716656 200
};
\addplot [draw=black, draw=black]
table{%
x  y
1 23.3360536819614
1.92208315076056 25.875
2.60674367430272 34.1666666666667
3.60825074500466 42.4583333333333
4.84121450662058 50.75
5.125 57.7494994356076
5.19801656233135 58.459816906599
};
\addplot [draw=black, draw=black]
table{%
x  y
8.06331823657427 75.7726789541032
9.16764910953434 83.9166666666667
8.75845049175434 92.2083333333333
9.25 95.2342270315962
10.7486063023193 100.5
11.3160633637239 108.791666666667
12.4611977227898 117.083333333333
13.1513942205873 125.375
13.3072373044449 133.666666666667
13.375 133.997054809016
15.2393192134062 141.958333333333
16.1225958047423 150.25
16.6648867796215 158.541666666667
17.5 163.077192331277
18.4730081161634 166.833333333333
19.1110372832179 175.125
19.8128332608396 183.416666666667
21.625 188.801344549611
22.4103400278666 191.708333333333
21.625 195.319332632706
20.8477996268777 200
};
\addplot [draw=black, draw=black]
table{%
x  y
1 38.9718676108949
1.41317787074965 42.4583333333333
2.73877299199213 50.75
2.88902632492796 59.0416666666667
3.54981592782777 67.3333333333333
4.05747570824601 75.625
4.53716015616346 82.6670127577829
};
\addplot [draw=black, draw=black]
table{%
x  y
6.28527021960362 101.00576679105
7.0104621354552 108.791666666667
7.32651960165779 117.083333333333
7.30676626840534 125.375
8.11421431523763 133.666666666667
9.06513104288054 141.958333333333
9.00996262273929 150.25
9.25 151.237979191966
10.968795048534 158.541666666667
10.7385809262074 166.833333333333
11.599058743861 175.125
11.5155121357376 183.416666666667
13.375 191.101109449293
13.5656339135066 191.708333333333
13.375 192.835707828401
12.4622628316648 200
};
\addplot [draw=black, draw=black]
table{%
x  y
1 65.6952238506469
1.10602379958708 67.3333333333333
1.6415184800205 75.625
2.06367242181055 83.9166666666667
2.62809455440616 92.2083333333333
2.79442821414748 100.5
3.25234570822683 107.481645307492
};
\addplot [draw=black, draw=black]
table{%
x  y
4.13703709347931 126.372604332843
4.16817301809213 133.666666666667
4.81531952699697 141.958333333333
4.60484693806855 150.25
5.125 157.242151911583
5.35466488566664 158.541666666667
6.2881408233167 166.833333333333
5.41505665969177 175.125
6.31309906763156 183.416666666667
7.12804419144465 191.708333333333
6.64066123701632 200
};
\addplot [draw=black, draw=black]
table{%
x  y
1 121.671627128727
1.72801222677014 125.375
1 131.919517439204
};
\addplot [draw=black, draw=black]
table{%
x  y
1 137.196854825156
1.37502307202933 141.958333333333
2.02391854310797 150.25
2.03905411532328 157.58099909393
};
\addplot [draw=black, draw=black]
table{%
x  y
2.03428639465361 176.121566097113
2.39862584971779 183.416666666667
2.71228460471586 191.708333333333
2.17361698005483 200
};
\draw (axis cs:13.375,42.6147843891443) node[
  scale=0.7,
  text=black,
  rotate=40.6
]{$\mathdefault{1.1}$};
\draw (axis cs:9.25,54.5514036899737) node[
  scale=0.7,
  text=black,
  rotate=52.2
]{$\mathdefault{1.2}$};
\draw (axis cs:6.47518433920807,67.3333333333333) node[
  scale=0.7,
  text=black,
  rotate=64.0
]{$\mathdefault{1.3}$};
\draw (axis cs:4.96529879225864,92.2083333333333) node[
  scale=0.7,
  text=black,
  rotate=75.0
]{$\mathdefault{1.4}$};
\draw (axis cs:3.3139429040692,117.083333333333) node[
  scale=0.7,
  text=black,
  rotate=82.2
]{$\mathdefault{1.5}$};
\draw (axis cs:1.17739942079313,166.833333333333) node[
  scale=0.7,
  text=black,
  rotate=270.5
]{$\mathdefault{1.6}$};
\end{axis}

\end{tikzpicture}}
    \caption{Price of delayed updates for different values of cost $c$ and incentive weight $\alpha$.}
    \label{fig:podu}
\end{figure}

In Fig.~\ref{fig:podu} we report the values of the price of delayed updates (PoDU) for different cost $c$ and incentive weight $\alpha$. From visual inspection of the plot, the presence of a large portion of the parameter space in which the PoDU is below $1.1$ is evident. This indicates that the efficiency loss achieved by the distributed solution is below $10\%$.
Interestingly, the zone where efficiency losses exceed $20\%$ is restricted for smaller values of the cost parameter $c$ and larger values of the incentive weight $\alpha$. This indicates that the incentive component primarily influences the selection of equilibrium paths. Additional numerical analysis, excluded here in the interest of conciseness, reveals that the PoDU increases logarithmically as the disparity between $c$ and $\alpha$ widens.
Another interesting remark is that the incentive for saving tokens is primarily responsible for the overall system efficiency and can only be approximately twice the value of the cost factor for the distributed system to be efficient.

\section{Conclusions and Future Works}\label{sec:conclusions}
In this paper, we analyzed the distributed and uncoordinated interaction of two sensors with the common goal of updating the receiver. We first presented the mathematical analysis of a cooperative static game with an incentive mechanism where the global objective is the minimization of the AoI, representing the freshness of the data. We found four NEs, corresponding to situations where none of the sensors transmits (equilibrium 1), either one updates the receiver (equilibria 2 and 3), and a mixed strategies NE where the two sensors transmit in a stochastic way (equilibrium 4). These equilibria arise depending on a threshold value, identified analytically. A repeated version of the game was also simulated over a finite time horizon, gaining insights into the prolonged interaction of the two sensors, particularly thanks to the definition of a new metric, the price of delayed updates (PoDU), from which we were able to identify a space for the ratio of system parameters that keeps the distributed solution performance close to the centralized (and unknown) optimum.
A possible direction for future research is extending this analysis to networks with more than two sensors. Additionally, studying partially correlated updates could provide deeper insights into AoI dynamics in multi-sender networks.

\balance

\bibliographystyle{IEEEtran}
\bibliography{IEEEabrv,AoI}

\end{document}